\documentclass[12pt]{article} 
\RequirePackage[colorlinks,citecolor=blue,urlcolor=blue]{hyperref}
\usepackage[utf8]{inputenc} 

\usepackage{geometry} 
\usepackage{hyperref}

\usepackage{graphicx} 

\usepackage{booktabs} 
\usepackage{array} 
\usepackage{paralist} 
\usepackage{verbatim} 
\usepackage{subfig} 
\usepackage{amsbsy}
\usepackage{amsmath}
\usepackage{amssymb}
\usepackage{amsfonts}
\usepackage{amscd}
\usepackage{amsthm}
\usepackage{cite}
\usepackage{float}
\usepackage[toc,page]{appendix}
\usepackage{natbib}
\usepackage{lineno}
\usepackage{mathrsfs}
\usepackage{bbm}
\usepackage{lineno}
\usepackage{todonotes}
\usepackage{setspace}
\usepackage{multirow}
\usepackage{fancyhdr} 
\usepackage{sectsty}
\allsectionsfont{\sffamily\mdseries\upshape} 

\usepackage[nottoc,notlof,notlot]{tocbibind} 
\usepackage[titles,subfigure]{tocloft} 

\graphicspath{{Figures/}}

\def\diffvar{\boldsymbol{\Sigma}}

\newcommand{\cDensity}[2]{\ensuremath{p(#1 \,|\,#2)}}
\newcommand{\density}[1]{\ensuremath{p(#1 )}}

\newcommand{\trOperator}[1]{\mbox{tr}\hspace{-0.1em}\left[ #1 \right]}

\newcommand{\order}[1]{{\cal O}\hspace{-0.1em}\left( #1 \right)}
\newcommand{\specialInverse}{^-}
\newcommand{\inverse}{^{-1}}
\newcommand{\transpose}{^t}

\newcommand{\momentum}{\boldsymbol{\rho}}

\newcommand{\rootTraitMean}{\boldsymbol{\nu}_0}
\newcommand{\rootSampleSize}{\kappa_0}
\newcommand{\parent}[1]{\text{pa}(#1)}

\newcommand{\nodePrecision}[1]{\mathbf{P}_{#1}}
\newcommand{\nodeVariance}[1]{\mathbf{\mathbf{V}}_{#1}}
\newcommand{\nodeVarianceFunction}[1]{\nodeVariance{}\hspace{-0.25em}\left( #1 \right)}
\newcommand{\paramList}{\boldsymbol{\theta}}

\newcommand{\general}[1]{g{}\hspace{-0.1em}\left( #1 \right)}

\newcommand{\nodeMean}[1]{\mathbf{m}_{#1}}
\newcommand{\nodeIndexOne}{i}
\newcommand{\nodeIndexTwo}{j}
\newcommand{\nodeIndexThree}{k}
\newcommand{\branchLength}[1]{t_{#1}}
\newcommand{\diffusionVariance}{\diffvar}

\newcommand{\traitIndex}{p}

\newcommand{\identityMatrix}[1]{\mathbf{I}_{#1}}

\newcommand{\nodePreMean}[1]{\mathbf{n}_{#1}}

\newcommand{\nodePrePrecision}[1]{\mathbf{Q}_{#1}}

\newcommand{\dd}{\text{d}}

\newcommand{\momentumAt}[1]{\rho_{#1}}
\newcommand{\branchRateSD}{s}

\newcommand{\observedMatrix}[1]{\boldsymbol{\delta}_{#1}}

\newcommand{\massMatrix}{\mathbf{M}}

\newcommand{\phylogeny}{\mathscr{F}}

\newcommand{\latentData}{\mathbf{Y}}

\newcommand{\nTaxa}{N}
\newcommand{\nTraits}{P}

\newcommand{\obs}{^{ \text{o} }}
\newcommand{\unobs}{^{\text{u}}}
\newcommand{\obsUnobs}{^{\text{uo}}}

\newcommand{\mvnDensity}[2]{\text{MVN}\hspace{-0.1em}\left(#1, #2\right)}

\newcommand{\gradient}[1]{\frac{\partial}{\partial #1}}

\newcommand{\fullGradient}{\nabla_{\allRates}}

\newcommand{\sequence}{\mathbf{S}}

\newcommand{\rate}[1]{\phi_{#1}}
\newcommand{\allRates}{\boldsymbol{\phi}}

\newcommand{\fullConditionalMean}[1]{\boldsymbol{\mu}_{#1}}
\newcommand{\fullConditionalVariance}[1]{\mathbf{Z}_{#1}}
\newcommand{\below}[1]{\lfloor #1 \rfloor}
\renewcommand{\above}[1]{\lceil #1 \rceil}
\newcommand{\expectation}[1]{\mathbb{E}\hspace{-0.1em}\left[ #1 \right]}

\def\newshortcut#1#2{%
\let#1=\undefined
\newcommand{#1}{#2}}

\newshortcut{\pyabove}{\cDensity{\latentData_{\nodeIndexOne}}{\latentData_{\above{\nodeIndexOne}}}}

\usepackage{titlesec}
\titleformat{\section}[block]{\sc\filcenter}{}{1em}{}
\titleformat{\subsection}[hang]{\it\filcenter}{}{1em}{}
\titlespacing*{\section}{0pt}{1\baselineskip}{0.1\baselineskip}
\usepackage{fancyhdr}
\begin{document}

\begin{flushright}
Version dated: \today\\
\end{flushright}

\bigskip
\medskip
\begin{center}

\noindent{\Large \bf Relaxed Random Walks at Scale}
\bigskip

\noindent{\normalsize \sc
	Alexander A.~Fisher$^{1}$,
	Xiang Ji$^{1}$,
	Philippe Lemey$^{2}$,
    and Marc A.~Suchard$^{1,3,4}$}\\
\noindent {\small
  \it $^1$Department of Biomathematics, David Geffen School of Medicine at UCLA, University of California,
  Los Angeles, United States \\
  \it $^2$Department of Microbiology and Immunology, Rega Institute, KU Leuven, Leuven, Belgium \\
  \it $^3$Department of Biostatistics, Jonathan and Karin Fielding School of Public Health, University
  of California, Los Angeles, United States \\
  \it $^4$Department of Human Genetics, David Geffen School of Medicine at UCLA, Universtiy of California,
  Los Angeles, United States} \\

\end{center}
\medskip
\noindent{\bf Corresponding author:} Marc A.~Suchard, Departments of Biostatistics, Biomathematics, and Human Genetics,
University of California, Los Angeles, 695 Charles E.~Young Dr., South,
Los Angeles, CA 90095-7088, USA; E-mail: \url{msuchard@ucla.edu}

\vspace{1in}

\clearpage

\section{Abstract}
Relaxed random walk (RRW) models of trait evolution introduce branch-specific rate multipliers to modulate the variance of a standard Brownian diffusion process along a phylogeny and more accurately model overdispersed biological data.
Increased taxonomic sampling challenges inference under RRWs as the number of unknown parameters grows with the number of taxa.
To solve this problem, we present a scalable method to efficiently fit RRWs and infer this branch-specific variation in a Bayesian framework.
We develop a Hamiltonian Monte Carlo (HMC) sampler to approximate the high-dimensional, correlated posterior that exploits a closed-form evaluation of the gradient of the trait data log-likelihood with respect to all branch-rate multipliers simultaneously.
Our gradient calculation achieves computational complexity that scales only linearly with the number of taxa under study. 
We compare the efficiency of our HMC sampler to the previously standard univariable Metropolis-Hastings approach while studying the spatial emergence of the West Nile virus in North America in the early 2000s.
Our method achieves an over 300-fold speed-increase over the univariable approach.  
Additionally, we demonstrate the scalability of our method by applying the RRW to study the correlation between five mammalian life history traits in a phylogenetic tree with $3650$ tips.

\section{Introduction}
Phylogenetic comparative methods are an indispensable tool to study the evolution of biological traits across taxa while controlling for their shared evolutionary history that confounds the inference of trait correlation \citep{felsenstein1985}.
Modern comparative methods usually entertain continuous, multivariate traits, although extensions to mixed discrete and continuous outcomes are readily available \citep{ives2009phylogenetic, cybis2015}.
Approaches typically model trait evolution as a Brownian diffusion or ``random walk" process that acts conditionally independently along the branches of a known or random phylogeny.
Specifically, the observed or unobserved (latent) trait value of a node in a phylogeny arises from a multivariate normal distribution centered on the latent trait value of its ancestral node with variance proportional to the units of time between nodes.
A strict Brownian diffusion model, however, is unable to accommodate the overdispersion in trait data that often emerges from real biological processes \citep{schluter1997likelihood}.
One such example arises when examining the dispersal rate of measurably evolving viral pathogens \citep{biek2007high}.
For example, if birds serve as the viral host, migratory patterns may induce inhomogeneous dispersal rates over time \citep{pybus2012unifying}.
In such cases, a strict Brownian diffusion model fails to capture, and therefore can also fail to predict, the spatial dynamics of an emerging epidemic. 

\cite{lemey2010} relax the strict Brownian diffusion assumption by introducing branch-rate multipliers that scale the variance of the Brownian diffusion process along each branch of the phylogeny.
This ``relaxed random walk" (RRW) model requires estimating $2\nTaxa - 2$ correlated branch-rate multipliers, where $\nTaxa$ is the number of taxa in the phylogeny.
\cite{lemey2010} take a Bayesian approach to parameter estimation where they infer the posterior distribution of the branch-rate multipliers via Markov chain Monte Carlo (MCMC) employing a simple univariable Metropolis-Hastings (UMH) proposal distribution \citep{hastings1970monte}. 
Since the rates remain correlated in the posterior, a random-scan \citep{liu2008monte} of UMH proposals inefficiently explores branch-rate space.
Specifically, univariable samplers force accepted proposals to be very close together to avoid a large number of rejection steps in the Markov chain simulation.
This results in high correlation between MCMC samples from the posterior, making point estimates of the branch-rate multipliers unreliable and slow to converge.
For example, in the mammalian life history study we explore here, a UMH sampler requires approximately 33 days of actual run time to achieve reasonable posterior estimates of branch-rate multipliers.
Despite this present drawback, RRWs find many impactful applications, e.g., in phylodynamics and phylogeography \citep{bedford2014integrating, faria2014early}.

To ameliorate the difficulties that high dimensional MCMC sampling presents, we propose adopting a geometry-informed sampling approach using Hamiltonian Monte Carlo (HMC).
HMC equates sampling from a probability distribution with simulating the trajectory of a puck sliding across a frictionless surface warped by the shape of the distribution \citep{neal2011mcmc}.
To map from this statistical problem to the physical one, we view the MCMC samples of our branch-rate multipliers as the ``position" of the puck and, then, for each positional dimension we introduce an associated momentum variable.
In this way, we extend a D-dimensional parameter space to 2D-dimensional phase space \citep{betancourt2017conceptual} and traverse the 2D phase space via differentiating the Hamiltonian and using a numerical integration method to offer proposal states for our MCMC chain.
The major limitation to HMC is calculating the gradient of the log-posterior with respect to all position parameters simultaneously.
Previous approaches for calculating gradients on  phylogenies have employed ``pruning"-type algorithms \citep{felsenstein1981} that scale quadratically with the number of taxa in the tree \citep{bryant2005likelihood}. Likewise, numerical approaches also scale quadratically.

In this paper, we derive a method to calculate the gradient with computational complexity that scales only linearly with the number of taxa.
We implement our method in the BEAST software package \citep{suchard2018}, a popular tool for the study and reconstruction of rooted, time-measured phylogenies.
We demonstrate the speed and accuracy of our linear-order gradient HMC versus previous best practices by examining the spread of the West Nile virus across the Americas in the early 2000s.
Finally, we use our technique to apply the RRW model to study the sensitivity of correlation estimates to model misspecification between mammalian adult body mass, litter size, gestation length, weaning age and litter frequency across $3650$ mammals, thereby demonstrating the scalability of our HMC implementation to tackling a previously intractable problem.

\section{Methods}
\subsection{Model and Inference} \label{sec:modeling}
Consider a known or random phylogeny $\phylogeny$ with $\nTaxa$ sampled tip nodes and $\nTaxa - 1$ internal and root nodes, each with an observed or latent continuous trait value $\latentData_{\nodeIndexOne}\, \in \, \mathbb{R}^{\nTraits}$.
To traverse the phylogeny $\phylogeny$, let node~$\parent{\nodeIndexOne}$ index the parent of node~$\nodeIndexOne$ with branch length $\branchLength{\nodeIndexOne}$ connecting the two nodes.
Then under the RRW model,
\begin{equation}
\latentData_{\nodeIndexOne} \sim \mvnDensity{\latentData_{\parent{\nodeIndexOne}}}{
\branchLength{\nodeIndexOne} \nodeVarianceFunction{\rate{\nodeIndexOne}}
} ,
\end{equation}
where the $\nTraits \times \nTraits$ matrix-valued function $\nodeVarianceFunction{\rate{\nodeIndexOne}}$ characterizes the branch-specific multivariate normal (MVN) increment that defines the diffusion process. We parameterize this function in terms of an unknown $\nTraits \times \nTraits$ positive-definite matrix $\diffusionVariance$ that describes the covariation between trait dimensions after controlling for shared evolutionary history and an unknown branch-rate multiplier $\rate{\nodeIndexOne}$.
Typical choices include
\begin{equation}
\nodeVarianceFunction{\rate{\nodeIndexOne}} =
\left\{
\begin{array}{rl}
\rate{\nodeIndexOne} \diffusionVariance & \text{rate-scalar parameterization}, \rate{\nodeIndexOne} > 0, \\
\frac{1}{\rate{\nodeIndexOne}} \diffusionVariance & \text{scale-mixture-of-normals parameterization}, \rate{\nodeIndexOne} > 0,  \\
e^{\rate{\nodeIndexOne}} \diffusionVariance & \text{unconstrained parameterization},\rate{\nodeIndexOne} \in \mathbb{R}, \ \text{and} \\
\diffusionVariance & \text{standard Brownian diffusion}.
\end{array}
\right.
\label{eq:varianceChoices}
\end{equation}
To complete the RRW model specification, we adopt a prior density on the unobserved trait at the parentless root node,
\begin{equation}
\density{\latentData_{2 \nTaxa - 1}} = \mvnDensity{\rootTraitMean}{
	\rootSampleSize^{-1} \diffusionVariance
	}
\end{equation}
with prior mean $\rootTraitMean$ and sample-size $\rootSampleSize$.

Letting $\allRates = (\rate{1}, \ldots, \rate{2\nTaxa - 2})$ and the observed data $\latentData = (\latentData_1, \ldots, \latentData_{\nTaxa})$ at the tips, we are interested in learning about the posterior
\begin{equation}
	\cDensity{\allRates, \diffusionVariance, \branchRateSD, \phylogeny, \paramList}{\latentData, \sequence}
	\propto
	\underbrace{
	\cDensity{\latentData}{\allRates, \diffusionVariance, \phylogeny}
	\cDensity{\sequence}{\phylogeny, \paramList}}_{\text{likelihood}}
	\underbrace{
	\cDensity{\allRates}{\branchRateSD}
	\density{\branchRateSD}
	\density{\diffusionVariance}
	\density{\phylogeny, \paramList}}_{\text{priors}},
	\label{eq:fullPosterior}
	\end{equation}
where $\branchRateSD$ is an unknown parameter characterizing our prior on $\allRates$ and $\paramList$ represents parameters of a molecular sequence substitution model for the evolution of aligned molecular sequence data $\sequence$.
Note that we follow usual convention \citep{cybis2015} and assume that $\latentData$ and $\sequence$ are conditionally independent given $\phylogeny$.
We follow the example of \cite{lemey2010} and place a log-normal prior distribution on $\allRates$ with mean $1$ and standard deviation $\branchRateSD$.
We further assume a relatively uninformative exponential prior on $\branchRateSD$ with mean $10$.
Additionally, we assign a Wishart conjugate prior with scale matrix $\identityMatrix{\nTraits}$ and $\nTraits$ degrees of freedom to $\diffusionVariance^{-1}$.

We use MCMC integration to approximate this posterior using a random-scan Metropolis-within-Gibbs approach \citep{levine2006, liu2008monte}.
One cycle of this scheme consists of sampling $\allRates, \diffusionVariance, \branchRateSD$ and then $(\phylogeny, \paramList)$ via

\begin{subequations}
\begin{align}
\label{eq:gibbsUpdates1}
\cDensity{\allRates}{\diffusionVariance, \branchRateSD, \phylogeny, \paramList, \latentData, \sequence}
&\propto
\cDensity{\latentData}{\allRates, \diffusionVariance, \phylogeny} \cDensity{\allRates}{\branchRateSD},\\
\label{eq:gibbsUpdates2}
\cDensity{\diffusionVariance}{\allRates, \branchRateSD, \phylogeny, \paramList, \latentData, \sequence}
&\propto
\cDensity{\latentData}{\allRates, \diffusionVariance, \phylogeny} \density{\diffusionVariance},\\
\label{eq:gibbsUpdates4}
\cDensity{\branchRateSD}{\allRates, \diffusionVariance, \phylogeny, \paramList, \latentData, \sequence}
&\propto
\cDensity{\allRates}{\branchRateSD} \density{\branchRateSD}, \ \text{and} \\
\label{eq:gibbsUpdates3}
\cDensity{\phylogeny, \paramList}{\allRates, \diffusionVariance, \branchRateSD, \latentData, \sequence}
&\propto
\cDensity{\latentData}{\allRates, \diffusionVariance, \phylogeny} \cDensity{\sequence}{\phylogeny, \paramList} \density{\phylogeny, \paramList},
\end{align}
\end{subequations}
where update (\ref{eq:gibbsUpdates3}) is unecessary when $\phylogeny$ is fixed, otherwise
efficient sampling from the density $\cDensity{\phylogeny, \paramList}{\allRates, \diffusionVariance, \branchRateSD, \latentData, \sequence}$ is well described elsewhere, see for example \cite{suchard2018}.
Updates (\ref{eq:gibbsUpdates2}) and (\ref{eq:gibbsUpdates4}) are straightforward due to the conjugate priors chosen in our model.
We turn our focus to the remaining component of our scheme, namely sampling from $\cDensity{\allRates}{\diffusionVariance, \branchRateSD, \phylogeny, \paramList, \latentData, \sequence}$.




\subsection{Hamiltonian Monte Carlo}
We wish to sample $\allRates$ jointly to avoid potentially high autocorrelation in the resulting MCMC chain.
To this end, we propose using HMC and begin with a brief description of how HMC maps sampling from a probability distribution to simulating a physical system.
In classical mechanics, the Hamiltonian is the sum of the kinetic and potential energy in a closed system.
To build the connection, we introduce auxiliary momentum variable $\momentum = (\momentumAt{1}, \ldots, \momentumAt{2 \nTaxa - 2})$ and write our Hamiltonian,
\begin{equation}
H(\allRates, \momentum) =
\underbrace{ - \log \cDensity{\allRates}{\diffusionVariance, \branchRateSD, \phylogeny, \paramList, \latentData, \sequence} }_{\text{potential energy}}
+
\underbrace{ \frac{1}{2} \momentum \transpose \massMatrix \momentum }_{\text{kinetic energy}},
\label{eq:Hamiltonian}
\end{equation}
where the mass matrix $\massMatrix$ weights our momentum variables.
The canonical distribution from statistical mechanics relates the joint density of state variables $\allRates$ and $\momentum$ to the energy in a system via the relationship,
\begin{equation}
	\cDensity{\allRates, \momentum}{\diffusionVariance, \branchRateSD, \phylogeny, \paramList, \latentData, \sequence} \propto e^{-H(\allRates, \momentum)}.
\label{eq:canonicalDistribution}
\end{equation}

Substituting our Hamiltonian into (\ref{eq:canonicalDistribution}), we observe that $\allRates$ and $\momentum$ are independent and recognize the marginal density of ${\momentum}$ to be MVN.
To start the HMC algorithm, we first sample $\momentum$ from this marginal density.
Then by differentiating $H(\allRates,\momentum)$, we generate Hamilton's equations of motion,
\begin{equation}
\begin{aligned}[c]
\frac{d \rate{\nodeIndexOne}}{dt} &= + \frac{\partial H}{\partial \rho_i}, \hspace{.2em} \text{and} \\
\frac{d \rho_i}{dt} &= - \frac{\partial H}{\partial \rate{\nodeIndexOne}} = \frac{\partial}{\partial \rate{\nodeIndexOne}} \log \cDensity{\allRates}{\diffusionVariance, \branchRateSD, \phylogeny, \paramList, \latentData, \sequence} \ \
\text{for all} \ \nodeIndexOne = 1, \dots, 2\nTaxa - 2 .
\end{aligned}
\end{equation}
We can use the resulting vector field in conjunction with a variety of numerical integration techniques to propose new states of $\allRates$ for our MCMC chain.
Consistent with typical construction \citep{neal2011mcmc}, we use the leapfrog method for numerical integration, where we follow the trajectory of $\momentum$ for a half-step before updating $\allRates$.
For a full discussion of HMC, see \cite{neal2011mcmc}.
Importantly, Hamilton's equations elicit a need to calculate $\fullGradient \log \cDensity{\allRates}{\diffusionVariance, \branchRateSD, \phylogeny, \paramList, \latentData, \sequence} = \left(\gradient{\rate{1}}, \dots, \gradient{\rate{2 \nTaxa - 2}}\right)^{\transpose} \log \cDensity{\allRates}{\diffusionVariance, \branchRateSD, \phylogeny, \paramList, \latentData, \sequence}$ to traverse phase space.

\subsection{Gradient of trait data log-likelihood}
\begin{figure}[H]
	\begin{center}
	\includegraphics[width=.4\textwidth]{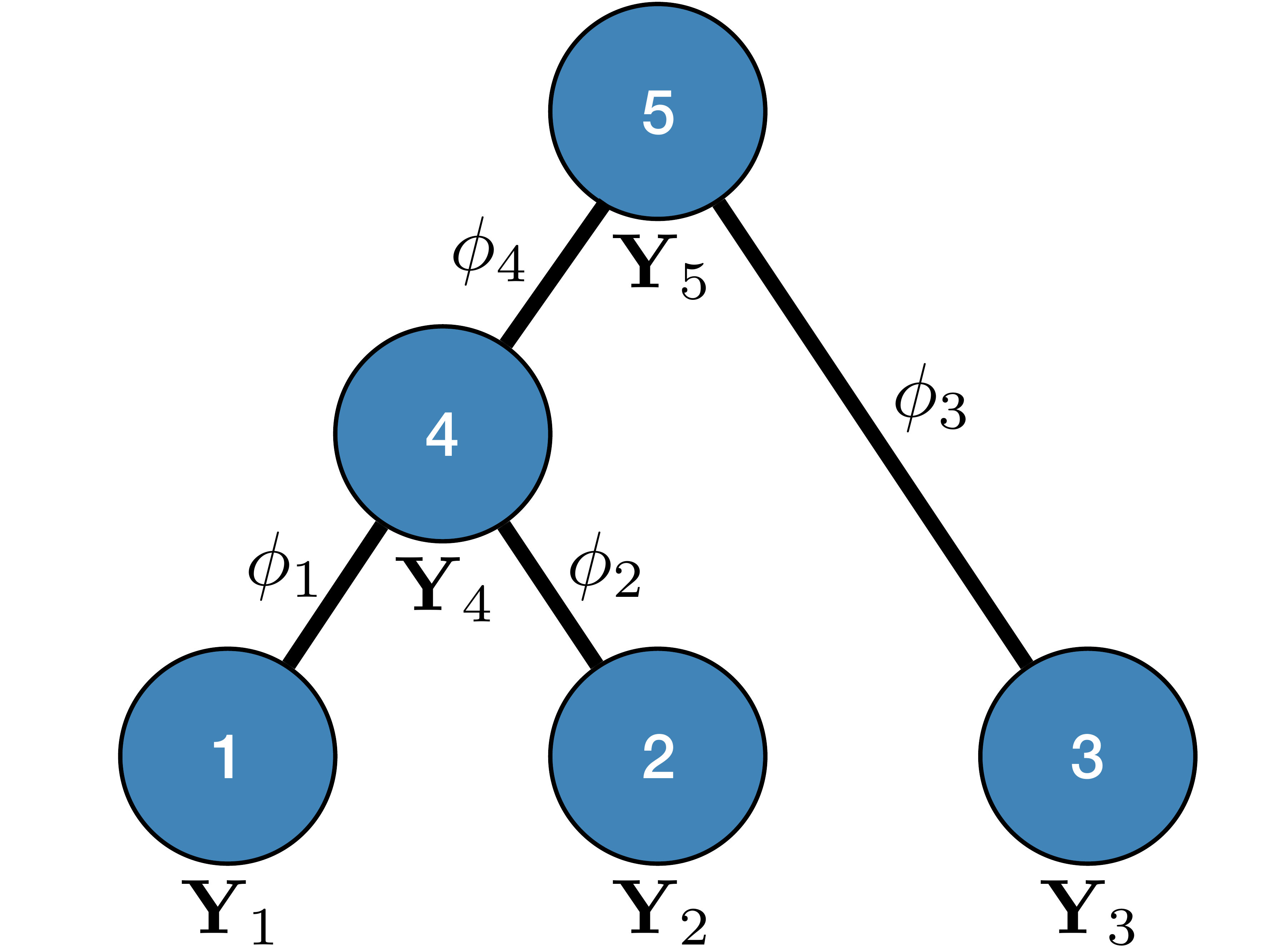}
	\end{center}
	\caption{Example tree with $\nTaxa = 3$ tips.
	Assume trait data $\latentData_{\nodeIndexOne}$ are fully observed for $\nodeIndexOne = \{1,2,3\}$.
	We write $\latentData_{\below{4}}$ and $\latentData_{\above{4}}$ to denote the observed data below and above node $4$ respectively.
	Specifically, $\latentData_{\below{4}} = \{\latentData_{1}, \latentData_2\}$ while  $\latentData_{\above{4}} = \{\latentData_{3}\}$.
	Partial likelihoods $\cDensity{\latentData_{\below{4}}}{\latentData_{4}}
	=
	\cDensity{\latentData_{1}, \latentData_{2}}{{\latentData_{4}}}$ and
	$\cDensity{\latentData_{4}}{\latentData_{\above{4}}}
	=
	\cDensity{\latentData_{4}}{\latentData_3}$. }
	\label{fig:treeTraversal}
	\end{figure}

A practical HMC sampler demands efficient calculation of $\fullGradient \log \cDensity{\allRates}{\diffusionVariance, \branchRateSD, \phylogeny, \paramList, \latentData, \sequence}$.
Differentiating the logarithm of (\ref{eq:gibbsUpdates1}), we obtain
\begin{equation}
\gradient{\rate{\nodeIndexOne}} \log \cDensity{\allRates}{\diffusionVariance, \branchRateSD, \phylogeny, \latentData, \sequence} = \gradient{\rate{\nodeIndexOne}} \log \cDensity{\latentData}{\allRates, \diffusionVariance, \phylogeny} + \gradient{\rate{\nodeIndexOne}} \log \cDensity{\allRates}{\branchRateSD}.
\label{eq:gradientPosterior}
\end{equation}
Our log-normal prior choice for $\allRates$ renders
evaluating the second term in Equation (\ref{eq:gradientPosterior}) trivial.
Here we develop a general recursive algorithm for calculating $\fullGradient \log \cDensity{\latentData}{\allRates, \diffusionVariance, \phylogeny}$.
To facilitate this development, consider splitting $\latentData$ into two disjoint sets relative to any node $\nodeIndexOne$ in $\phylogeny$.
We define $\latentData_{\below{\nodeIndexOne}}$ as the observed data descendant of node $\nodeIndexOne$ and $\latentData_{\above{\nodeIndexOne}}$ as the observed data ``above" (or not descendent of) node $\nodeIndexOne$.
For clarity, see Figure (\ref{fig:treeTraversal}).
In the following, we drop the dependence of the log-likelihood on $\allRates$, $\diffusionVariance$ and $\phylogeny$ for notational convenience. To begin,

\begin{equation}\begin{aligned}
\gradient{\rate{\nodeIndexOne}} \left[
\log \density{\latentData}
\right]
= & \,\, \gradient{\rate{\nodeIndexOne}} \left[ \density{\latentData} \right]
\Big/  \density{\latentData} \\
= & \,\, \gradient{\rate{\nodeIndexOne}} \left[ \int
 \cDensity{\latentData_{\below{\nodeIndexOne}}}{\latentData_{\nodeIndexOne}}
 \cDensity{\latentData_{\nodeIndexOne}}{\latentData_{\above{\nodeIndexOne}}}
 \density{\latentData_{\above{\nodeIndexOne}}}
 \dd \latentData_{\nodeIndexOne} \right]
\Big/  \density{\latentData} \\
= & \,\, \int \gradient{\rate{\nodeIndexOne}} \left[
 \cDensity{\latentData_{\below{\nodeIndexOne}}}{\latentData_{\nodeIndexOne}}
 \cDensity{\latentData_{\nodeIndexOne}}{\latentData_{\above{\nodeIndexOne}}}
 \density{\latentData_{\above{\nodeIndexOne}}} \right]
 \dd \latentData_{\nodeIndexOne}
\Big/  \density{\latentData} \\
= & \,\, \int
 \cDensity{\latentData_{\below{\nodeIndexOne}}}{\latentData_{\nodeIndexOne}}
 \gradient{\rate{\nodeIndexOne}} \left[
 \cDensity{\latentData_{\nodeIndexOne}}{\latentData_{\above{\nodeIndexOne}}}
 \right]
 \density{\latentData_{\above{\nodeIndexOne}}}
 \dd \latentData_{\nodeIndexOne}
\Big/  \density{\latentData}.
%
\label{eq:differential}
\end{aligned}\end{equation}
The last equality above follows from the fact that $\rate{\nodeIndexOne}$ is associated only with the branch above node $\nodeIndexOne$.
Therefore when we condition on $\latentData_{\nodeIndexOne}$, $\latentData_{\below{\nodeIndexOne}}$ is independent of $\rate{\nodeIndexOne}$.
Similarly, $\latentData_{\above{\nodeIndexOne}}$ evolves independent of $\rate{\nodeIndexOne}$.
To proceed with the differential above, we use the fact that $\cDensity{\latentData_{\nodeIndexOne}}{\latentData_{\above{\nodeIndexOne}}}$ follows a MVN distribution with as of yet undetermined mean $\nodePreMean{\nodeIndexOne}$ and precision $\nodePrePrecision{\nodeIndexOne}$ (see Appendix (\nameref{sec:preOrderDerivation}) for a detailed derivation).
We extract the middle term from Equation (\ref{eq:differential}) and find
\begin{equation}\begin{aligned}
 \gradient{\rate{\nodeIndexOne}} \left[
 \cDensity{\latentData_{\nodeIndexOne}}{\latentData_{\above{\nodeIndexOne}}}
 \right] &=
 \frac{1}{2}
 \left\{
 	\left( \latentData_{\nodeIndexOne} - \nodePreMean{\nodeIndexOne} \right)\transpose \nodePrePrecision{\nodeIndexOne} \hspace{0.1em}  \branchLength{\nodeIndexOne}
 		 \gradient{\rate{\nodeIndexOne}} \left[ \nodeVarianceFunction{\rate{\nodeIndexOne}} \right]
 	\hspace{-0.1em} \nodePrePrecision{\nodeIndexOne} \left( \latentData_{\nodeIndexOne} - \nodePreMean{\nodeIndexOne} \right)
 \right.
 \\
 & \quad \quad \quad \quad \quad \quad
 \left.
 	-
 	\trOperator{
 		\nodePrePrecision{\nodeIndexOne} \hspace{0.1em}  \branchLength{\nodeIndexOne} \gradient{\rate{\nodeIndexOne}} \left[ \nodeVarianceFunction{\rate{\nodeIndexOne}} \right]
 	}
 \right\}
  \cDensity{\latentData_{\nodeIndexOne}}{\latentData_{\above{\nodeIndexOne}}} ,
\end{aligned}
\label{eq:differential_mvn}
\end{equation}
using the differential properties
\begin{equation}\begin{aligned}
\dd \nodePrePrecision{\nodeIndexOne} &= - \nodePrePrecision{\nodeIndexOne} \hspace{-0.1em} \left( \dd \nodePrePrecision{\nodeIndexOne}^{-1} \right) \nodePrePrecision{\nodeIndexOne} \ \text{ and} \\
\dd \hspace{-0.1em} \left| \nodePrePrecision{\nodeIndexOne}^{-1} \right| &= \left| \nodePrePrecision{\nodeIndexOne}^{-1} \right| \hspace{-0.1em} \trOperator{ \nodePrePrecision{\nodeIndexOne} \dd \nodePrePrecision{\nodeIndexOne}^{-1} },
\end{aligned}\end{equation}
found in, e.g., \cite{matrixcookbook}.

\newcommand{\functional}[1]{\mathbf{F}\hspace{-0.1em}\left( #1 \right)}
\newcommand{\nastyPrecision}[1]{\boldsymbol{\Upsilon}_{#1}}
\newcommand{\nastyDeterminant}[1]{\boldsymbol{\chi}_{#1}}

To simplify notation, we let function
\begin{equation}\begin{aligned}
\functional{\latentData_{\nodeIndexOne}} =
 \frac{1}{2}
 \left\{
 	\left( \latentData_{\nodeIndexOne} - \nodePreMean{\nodeIndexOne} \right)\transpose
 	\nastyPrecision{\nodeIndexOne}
 	\left( \latentData_{\nodeIndexOne} - \nodePreMean{\nodeIndexOne} \right)
 %
 	-
 	\trOperator{
		\nastyDeterminant{\nodeIndexOne}
 	}
 \right\} ,
 \label{eq:functional}
\end{aligned}\end{equation}
where
$\nastyPrecision{\nodeIndexOne} =
\nodePrePrecision{\nodeIndexOne} \hspace{0.1em}  \branchLength{\nodeIndexOne} \gradient{\rate{\nodeIndexOne}} \left[ \nodeVarianceFunction{\rate{\nodeIndexOne}} \right] \hspace{-0.1em} \nodePrePrecision{\nodeIndexOne}$ and
$\nastyDeterminant{\nodeIndexOne} =
\nodePrePrecision{\nodeIndexOne} \hspace{0.1em}  \branchLength{\nodeIndexOne} \gradient{\rate{\nodeIndexOne}} \left[ \nodeVarianceFunction{\rate{\nodeIndexOne}} \right]$.
Substituting Equation (\ref{eq:functional}) back into Equation (\ref{eq:differential}), we observe that
\begin{equation}\begin{aligned}
\gradient{\rate{\nodeIndexOne}} \left[
\log \density{\latentData}
\right]
= & \,\, \int
 \functional{\latentData_{\nodeIndexOne}}
 \cDensity{\latentData_{\below{\nodeIndexOne}}}{\latentData_{\nodeIndexOne}}
 \cDensity{\latentData_{\nodeIndexOne}}{\latentData_{\above{\nodeIndexOne}}}
 \density{\latentData_{\above{\nodeIndexOne}}}
 \dd \latentData_{\nodeIndexOne}
\Big/  \density{\latentData} \\
= & \,\, \int
 \functional{\latentData_{\nodeIndexOne}}
 \cDensity{\latentData_{\nodeIndexOne}}{\latentData}
 \dd \latentData_{\nodeIndexOne} \\
= & \,\, \expectation{ \functional{\latentData_{\nodeIndexOne}} \mid \latentData}.
\end{aligned}
\label{eq:expectationOfF}
\end{equation}
When $\latentData_{\nodeIndexOne}$ is fully observed (typically $\nodeIndexOne \leq N$), this expectation collapses to the direct evaluation of $\functional{\latentData_{\nodeIndexOne}}$.
When $\nodeIndexOne = \nTaxa + 1, \ldots, 2 \nTaxa - 2$
 or if $\latentData_{\nodeIndexOne}$ is partially observed for $\nodeIndexOne = 1,\ldots, \nTaxa$,
we require $\cDensity{\latentData_{\nodeIndexOne}}{\latentData}$.
From Bayes' theorem, $\cDensity{\latentData_{\nodeIndexOne}}{\latentData} \propto \cDensity{\latentData_{\below{\nodeIndexOne}}}{\latentData_{\nodeIndexOne}}
 \cDensity{\latentData_{\nodeIndexOne}}{\latentData_{\above{\nodeIndexOne}}}$.
Partial likelihood $\cDensity{\latentData_{\below{\nodeIndexOne}}}{\latentData_{\nodeIndexOne}}$ is proportional to a MVN density characterized by computable mean $\nodeMean{\nodeIndexOne}$ and precision $\nodePrecision{\nodeIndexOne}$ \citep{pybus2012unifying}.
Using this fact, $\cDensity{\latentData_{\nodeIndexOne}}{\latentData}$ becomes MVN with
mean
$
\fullConditionalMean{\nodeIndexOne} =
\fullConditionalVariance{\nodeIndexOne} \left(
	\nodePrecision{\nodeIndexOne} \nodeMean{\nodeIndexOne} + \nodePrePrecision{\nodeIndexOne} \nodePreMean{\nodeIndexOne}
\right)
$
and
variance
$
\fullConditionalVariance{\nodeIndexOne} =
\left[
	\nodePrecision{\nodeIndexOne} + \nodePrePrecision{\nodeIndexOne}
\right] \inverse.
$
 When tip $\nodeIndexOne$ is partially observed, we partition $\latentData_{\nodeIndexOne} = \left( \latentData_{\nodeIndexOne} \unobs, \ \latentData_{\nodeIndexOne} \obs \right) \transpose$ into its unobserved and observed entries.
Using properties of the conditional MVN,
$\cDensity{\latentData_{\nodeIndexOne}}{\latentData}$ becomes degenerate with mean
\begin{equation}
\fullConditionalMean{\nodeIndexOne} =
\begin{bmatrix}
\nodePreMean{\nodeIndexOne} \unobs - \left( \nodePrePrecision{\nodeIndexOne} \unobs \right) \inverse
\nodePrePrecision{\nodeIndexOne} \obsUnobs
\left(
	\latentData_{\nodeIndexOne} \obs - \nodePreMean{\nodeIndexOne} \obs
\right)
\\
\latentData_{\nodeIndexOne} \obs
\end{bmatrix}
\end{equation}
and variance
\begin{equation}
	\fullConditionalVariance{\nodeIndexOne} =
	\begin{bmatrix}
	\left(
	\nodePrePrecision{\nodeIndexOne} \unobs
	\right) \inverse
	& 0\\
	0 & 0
\end{bmatrix}.
\end{equation}
Finally, for both partially and completely unobserved cases above,
\begin{equation}\begin{aligned}
\expectation{ \functional{\latentData_{\nodeIndexOne}} \mid \latentData} &=
 \frac{1}{2}
 \left\{
 \trOperator{
 	\fullConditionalVariance{\nodeIndexOne} \nastyPrecision{\nodeIndexOne}
 }
 +
 \left( \fullConditionalMean{\nodeIndexOne} - \nodePreMean{\nodeIndexOne} \right)\transpose
 	\nastyPrecision{\nodeIndexOne}
 \left( \fullConditionalMean{\nodeIndexOne} - \nodePreMean{\nodeIndexOne} \right)
 	-
 	\trOperator{
		\nastyDeterminant{\nodeIndexOne}
 	}
 \right\}.
 \label{eq:linearGradient}
\end{aligned}\end{equation}
Equation (\ref{eq:linearGradient}) provides a recipe to compute $\fullGradient \log \cDensity{\latentData}{\allRates, \diffusionVariance, \phylogeny}$ using the means and precisions that characterize partial data likelihoods $\cDensity{\latentData_{\nodeIndexOne}}{\latentData_{\above{\nodeIndexOne}}}$ and $\cDensity{\latentData_{\below{\nodeIndexOne}}}{\latentData_{\nodeIndexOne}}$.

\subsection{Tree Traversals} \label{sec:algorithms}
We introduce post- and pre-order tree traversals to recursively calculate all partial data likelihood means and precisions in computational complexity $\order{\nTaxa}$ that scales linearly with $\nTaxa$.
To begin, let nodes $\nodeIndexOne$ and $\nodeIndexTwo$ be daughters of node $\nodeIndexThree$.
Following \cite{hassler2019inferring}, let $\observedMatrix{\nodeIndexThree} = \text{diag}(\delta_{\nodeIndexThree 1}, \ldots \delta_{\nodeIndexThree \nTraits})$ for $\nodeIndexThree = 1, \ldots, \nTaxa$ be a diagonal matrix with indicator elements $\delta_{\nodeIndexThree \traitIndex}$ that take value $1$ if $\text{Y}_{\nodeIndexThree \traitIndex}$ is observed and $0$ if not. For the post-order traversal,
\begin{equation}
\cDensity{\latentData_{\below{\nodeIndexOne}}}{\latentData_{\nodeIndexOne}} \propto \mvnDensity{\latentData_{\nodeIndexOne}; \ \nodeMean{\nodeIndexOne}}{\nodePrecision{\nodeIndexOne}},
\end{equation}
with post-order mean $\nodeMean{\nodeIndexOne}$ and precision $\nodePrecision{\nodeIndexOne}$.
For $\nodeIndexThree = 1, \ldots, 2 \nTaxa - 1$ in post-order, we build the precision via

\begin{equation}
\nodePrecision{\nodeIndexThree} =  \begin{cases} \infty \times \observedMatrix{\nodeIndexThree} \ \ \ \text{if k is a tip}\\
(\nodePrecision{\nodeIndexOne}^* + \nodePrecision{\nodeIndexTwo}^*)  \ \ \text{otherwise,}
\end{cases}
\end{equation}
with the definition that $\infty \times 0 = 0 $ and
\begin{equation}
\nodePrecision{\nodeIndexOne}^* = \Big( \nodePrecision{\nodeIndexOne} \specialInverse + \branchLength{\nodeIndexOne} \observedMatrix{\nodeIndexOne} \nodeVarianceFunction{\rate{\nodeIndexOne}} \observedMatrix{\nodeIndexOne} \Big) \specialInverse
\ \ \ \text{and} \ \ \
\nodePrecision{\nodeIndexTwo}^* = \Big( \nodePrecision{\nodeIndexTwo} \specialInverse + \branchLength{\nodeIndexTwo} \observedMatrix{\nodeIndexTwo} \nodeVarianceFunction{\rate{\nodeIndexTwo}} \observedMatrix{\nodeIndexTwo} \Big) \specialInverse,
\end{equation}
where the pseudo-inverse, defined and developed by \cite{bastide2018inference, hassler2019inferring}, is described in Appendix (\nameref{sec:pseudoInverseExplanation}).
At the tips, we build the mean
$
\nodeMean{\nodeIndexThree}
= \observedMatrix{\nodeIndexThree} \odot \latentData_{\nodeIndexThree}
$
where $\odot$ is the elementwise dot product,
and for the internal nodes, $\nodeMean{\nodeIndexThree}$ is a solution to
\begin{equation}
	\nodePrecision{\nodeIndexThree}
	\nodeMean{\nodeIndexThree}
	=
	 (\nodePrecision{\nodeIndexOne}^* \nodeMean{\nodeIndexOne} + \nodePrecision{\nodeIndexTwo}^* \nodeMean{\nodeIndexTwo}).
\end{equation}
For a proof of these post-order updates, see \cite{hassler2019inferring} (Supplemental Material).

To compute $\cDensity{\latentData_{\nodeIndexOne}}{\latentData_{\above{\nodeIndexOne}}}$, we traverse the tree in pre-order fashion according to our generalized version of the recursive algorithm proposed by \cite{cybis2015}.
See Appendix (\nameref{sec:preOrderDerivation}) for a derivation of our generalized pre-order update.
For the pre-order traversal,

\begin{equation}
\cDensity{\latentData_{\nodeIndexOne}}{\latentData_{\above{\nodeIndexOne}}} = \mvnDensity{\latentData_{\nodeIndexOne}; \ \nodePreMean{\nodeIndexOne}}{\nodePrePrecision{\nodeIndexOne}}.
\label{eq:preOrderDistribution}
\end{equation}
For $\nodeIndexOne = 2\nTaxa - 1, \ldots, 1$ looking down the tree, we update our pre-order precision,

\begin{equation}
\nodePrePrecision{\nodeIndexOne} =
\begin{cases}
	\rootSampleSize \diffusionVariance \inverse \ \ \ \
	\text{if i is root}\\
\Big( (\nodePrePrecision{\nodeIndexOne}^{*}) \inverse + t_i \nodeVarianceFunction{\rate{\nodeIndexOne}} \Big)^{-1} \ \ \ \ \text{otherwise}
\end{cases}
\end{equation}
at each node where
\begin{equation}
\nodePrePrecision{\nodeIndexOne}^{*} = \nodePrecision{\nodeIndexTwo}^* + \nodePrePrecision{\nodeIndexThree}.
\end{equation}
We also keep track of the pre-order mean at each node via
\begin{equation}
\nodePreMean{\nodeIndexOne} =
\begin{cases}
\rootTraitMean \ \ \ \
\text{if i is root}\\
(\nodePrePrecision{\nodeIndexOne}^{*}) \inverse \big( \nodePrecision{\nodeIndexTwo}^* \nodeMean{\nodeIndexTwo}
+
\nodePrePrecision{\nodeIndexThree} \nodePreMean{\nodeIndexThree} \big) \ \ \ \
\text{otherwise}.
\end{cases}
\end{equation}

Both traversals visit each node exactly once and perform a matrix inversion as their most costly operation, providing an $\order{\nTaxa \nTraits^3}$ algorithm.
However, as we observe in Equation (\ref{eq:varianceChoices}), generally $\nodeVarianceFunction{\rate{\nodeIndexOne}} = \general{\rate{\nodeIndexOne}} \diffusionVariance$.
In this case, we can further reduce the computational complexity to $\order{\nTaxa \nTraits ^2}$ by factoring out $\diffusionVariance$.
Instead of inverting $\nodeVarianceFunction{\rate{\nodeIndexOne}}$ at each step, we only need to invert $\diffusionVariance$ at most once per likelihood or gradient evaluation.
\section{Results} 
\subsection{West Nile Virus}
West Nile virus (WNV) is responsible for more than 1,500 deaths and caused over 700,000 illnesses since first reported in North America in 1999.
The virus typically spreads via mosquito bites; however, the primary host is birds.
First identified in New York City, WNV spread to the Pacific coast by 2003 and reached south into Argentina by 2005 \citep{petersen2013west}.
We examine whole aligned viral genomes (11,029 nt) and geographic data on $104$ cases of WNV collected between 1999 and 2007 \citep{pybus2012unifying}.
In cases where only the year of sampling is known, we set the sampling date to the midpoint of that year.
Previous authors have recorded latitude and longitude geographic sampling information by converting zip code locations using ZIPList5.
For 27 of the specimens, only the U.S. or Mexican state of discovery is known and so we have augmented sampling data with the coordinates of the centroid of the state \citep{pybus2012unifying}.

Here we study the simultaneous evolution and dispersal of WNV as it spreads across North America, following the modeling choices of \cite{pybus2012unifying}.
We define geographic location as our trait of interest $\latentData$ within a RRW and infer rates $\allRates$ using our new HMC method.
In two separate inference scenarios, we compare the computational efficiency of our method to the random-scan UMH approach employed by \cite{pybus2012unifying}.
\begin{figure}
	\begin{center}
	\includegraphics[width=.7\textwidth]{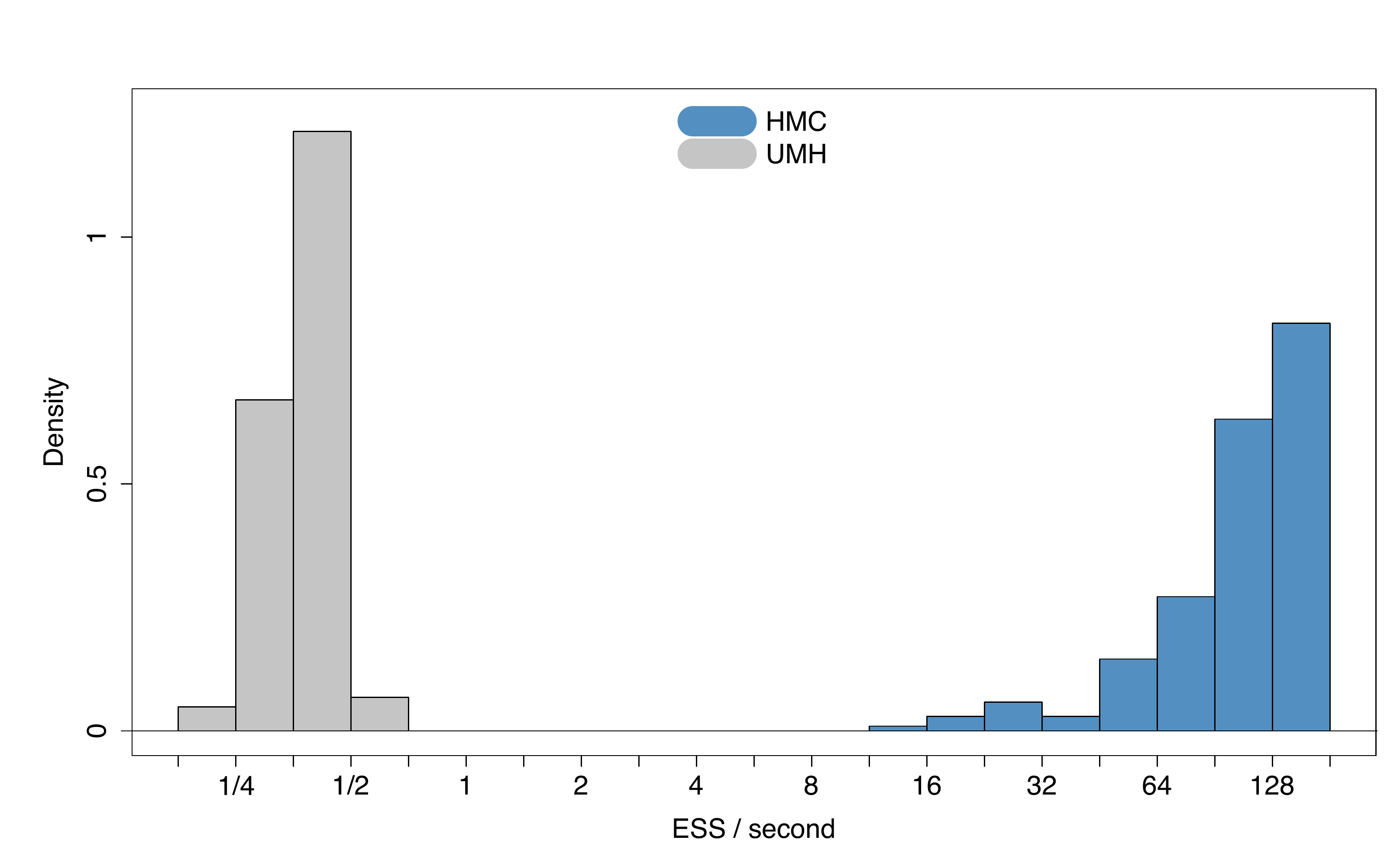}
	\end{center}
	\caption{Comparing computational efficiency of Hamiltonian Monte Carlo (HMC) and univariable Metropolis-Hastings (UMH) transition kernels through effective sample size (ESS) per unit time in West Nile virus (WNV) phylogeography.
	}
	\label{fig:ess}
	\end{figure}

To begin, we set up a RRW model with log-normal prior on rates $\allRates$ with mean $= 1$ and standard deviation $\branchRateSD$ and use a general time-reversible (GTR) + $\Gamma$ substitution model with a log-normal relaxed molecular clock.
We use the UMH transition kernel to run a 250 million state MCMC chain simulation to obtain  posterior mean estimates of $\diffusionVariance$ and $\branchRateSD$ as well as a maximum clade crediblility (MCC) tree.
In scenario (\textbf{a}) we use these fixed model parameters and topology to strictly sample $\allRates$ using both the HMC and UMH transition kernels.
Under this fixed analysis we run our HMC-based chain for 1 million states and a UMH-based chain for 150 million states.
We use effective sample size (ESS) of the  posterior $\rate{\nodeIndexOne}$ samples for all $\nodeIndexOne$ divided by computational runtime to evaluate the performance of each MCMC approach and report densities of ESS/second across all branches in Figure (\ref{fig:ess}).
ESS/second is averaged across five runs each with uniform (0-10) random initial branch-rate multipliers.
The median ESS/second across $\allRates$ is $118$ and $0.377$ for the HMC and UMH transition kernels respectively.
This demonstrates a $312$-fold speed increase.
Additionally, the minimum ESS/second is $12.4$ with HMC and $0.183$ with UMH, exhibiting a $67.7$-fold speed-up for the ``least well'' explored $\rate{\nodeIndexOne}$.
In scenario (\textbf{b}), we use a random starting tree and jointly estimate all parameters ($\allRates, \diffusionVariance$, $\branchRateSD$, $\phylogeny$ and $\paramList$) of the full posterior (\ref{eq:fullPosterior}).
Since branch-specific $\rate{\nodeIndexOne}$ are no longer identifiable when $\phylogeny$ is random, we compare square jump distance across all $\allRates$ between samples from the posterior under both MCMC regimes to compare efficiency.
We run HMC chains for 22.5 million states so that we are sampling from the posterior distribution of all parameters and we save the state of BEAST.
Subsequently, we run both HMC and UMH chains from the same saved states and compute lag-7 square jump distance to adjust for the relative weight of the transition kernel in the full analysis.
Since the UMH sampler updates only one branch-rate multiplier at a time, we compare square jump distance between samples of our HMC chain with samples from the UMH chain that are lagged $(2\nTaxa - 2)\times$ (i.e.~$206\times$) farther apart.
We run each MCMC simulation until we obtain 5000 samples from the posterior and report the average median across five separate runs.
In this comparison, we find that the average square jump distances from five separate runs
is $1457$ and $128.2$ for the HMC and UMH chains respectively.

\begin{figure}[H]
	\begin{center}
	\includegraphics[width=0.8\textwidth]{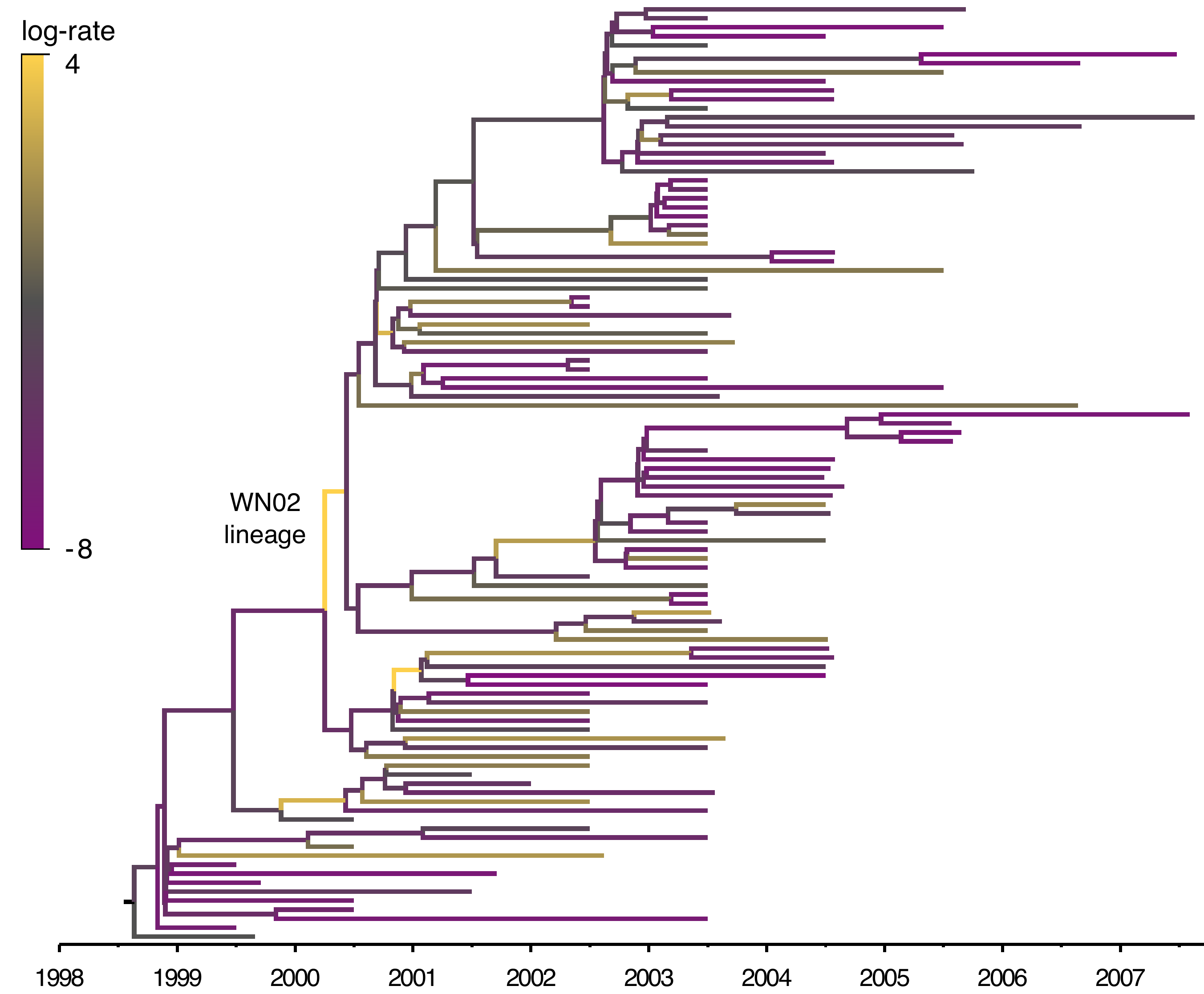}
	\end{center}
	\caption{Maximum clade crediblility (MCC) tree resulting from Hamiltonian Monte Carlo (HMC) inference under phylogeographic relaxed random walk (RRW) of West Nile virus.
	We color branches by posterior mean branch-rate parameters $\allRates$.}
	\label{fig:wnvTree}
	\end{figure}

In Figure (\ref{fig:wnvTree}) we report the MCC tree, 
obtained from applying HMC to the RRW model as described in scenario (\textbf{b}), where substitution rate variation is accounted for by the molecular clock model.
The branch with the highest posterior dispersal rate starts the WN02 lineage identified by \cite{gray2010evolutionary}.
The clade of New York isolates sampled in 1999, however, maintains a much slower dispersal rate. Sufficiently precise posterior estimates under the previous UMH approach remain unavailable in reasonable time to expose these differences.

\subsection{Mammalian Life History}
Life history theory aims to explain how traits such as adult body mass, litter size and lifespan evolve to optimize reproductive success \citep{stearns2000life}.
Life history theory finds important applications in determining a species' fecundity and predicting extinction risk in response to changing environmental stimuli \citep{pacifici2017species, fritz2009geographical, de2019demographic},
but due to the sparseness of much life history data, it is essential to understand how traits covary to make meaningful predictions \citep{santini2016trait}.
To determine which traits covary, comparative mammalian life-history studies posit a `fast–slow' continuum, claiming small mammals are typically `fast', characterized by early maturation, large litters and shorter lifespans, while larger mammals are typically `slow' and present contrasting characteristics \citep{oli2004fast, millar1983life}. Under this framework, certain traits such as gestation length, weaning age and body mass are predicted to be positively correlated, but reported estimates of positive correlation from data may be artifacts of the restrictive assumptions of strict Brownian diffusion modeling.
Here we re-evaluate this claim within the RRW of trait evolution made tractable through $\order{\nTaxa}$ HMC sampling.

\begin{figure}[H]
\begin{center}
\includegraphics[width=.8\textwidth]{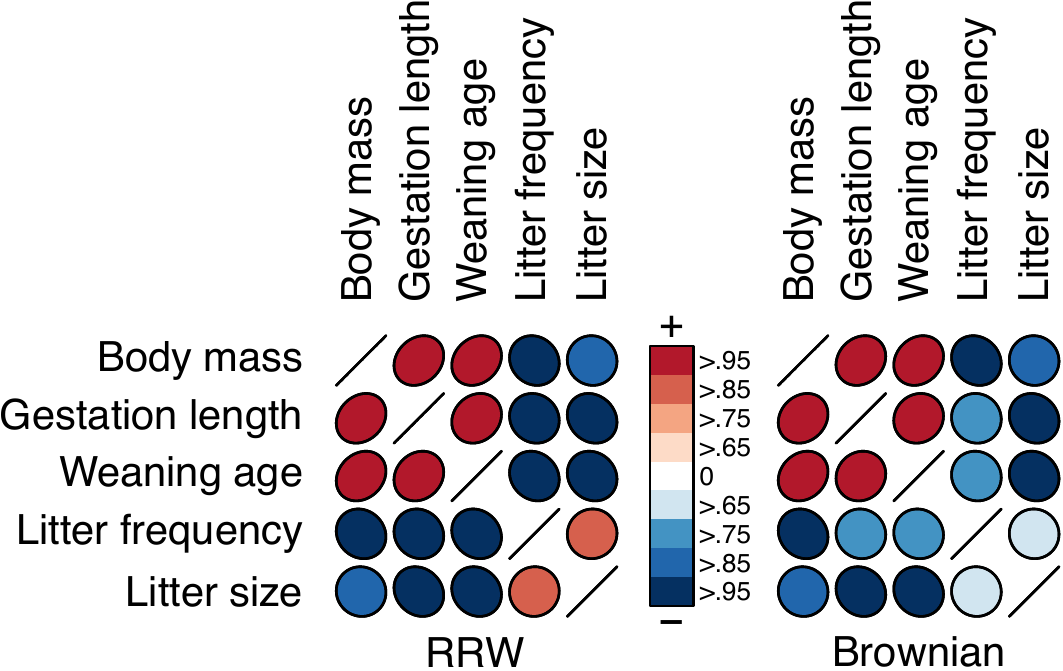}
\end{center}
\caption{Posterior mean correlation between mammalian life history traits under the RRW and strict Brownian diffusion model. Shape of ellipse indicates strength and sign of correlation, while colors indicate the posterior probability that the correlation is positive (red) or negative (blue).}
\label{fig:mammalCorrelation}
\end{figure}

Under the RRW, we infer correlation between five life history traits from the PanTHERIA data set \citep{jones2009pantheria}, namely body mass, litter size, gestation length, weaning age and litter frequency across $3650$ mammalian species related by the fixed supertree of \cite{fritz2009geographical}.
To obtain this subset of the supertree, we only consider taxa for which at least one of these five traits is observed.
We take the intersection of this set of taxa with those in the fixed supertree of \cite{fritz2009geographical} and prune all other observations from the tree.
We log-transform and standardize the trait measurements and subsequently estimate posterior mean correlations between each pair of traits under the RRW using an HMC-based chain for 300 thousand states.
To gauge the effect of a heterogeneous diffusion process on the correlation between traits, we also make inference using the strict Brownian diffusion model where the $\allRates$ are all identically $1$. Here we perform MCMC inference on the diffusion matrix $\diffusionVariance$ for 100 thousand states.
We report posterior mean estimates of correlation between each pair of traits under both the RRW and strict Brownian diffusion in Figure (\ref{fig:mammalCorrelation}).
In most cases, the RRW reassuringly confirms analysis under the more limited model. However, in some cases our confidence in the sign of the correlation differs between models and
in one instance the sign of the posterior mean correlation disagrees.
Under the RRW, we observe positive posterior mean correlation  of $0.021$ between litter frequency and litter size with posterior odds ratio $2.63$ that the correlation is positive.
Under the strict Brownian diffusion model we observe a negative posterior mean correlation of $-0.018$ with posterior odds ratio $2.25$ of being negative, indicating slightly weaker belief in the correlation's sign under the strict model.



%

\section{Discussion}

Previous MCMC techniques to investigate trait evolution under the RRW model scale poorly with large data sets.
Specifically the UMH transition kernel is ineffective for sampling correlated, high dimensional parameter space.
We provide a remedy by using an HMC transition kernel to sample all branch-rate multipliers simultaneously.
To improve the speed of HMC we derive an algorithm for calculating the gradient of the trait data log-likelihood.
This gradient calculation achieves $\order{\nTaxa}$ computational speed, a vast improvement compared to both numerical and pruning methods for calculating the gradient that typically require $\order{\nTaxa^2}$.

We observe over 300-fold speed-up when comparing our HMC transition kernel to UMH in the spread of the WNV across North America in the early 2000s.
The resulting MCC tree reveals that the largest dispersal rate precedes the most recent common ancestor of the WN02 lineage.
Subsequently, the dispersal rates slow down through the WN02 clade.
This suggests that this clade developed after some rapid geographic displacement.
Interestingly, the appearance of smaller branch-rate multipliers within the WN02 lineage is consistent with the slowing speed of sequence evolution as described in \cite{snapinn2007declining}.

As exhibited in Figure (\ref{fig:ess}), ESS from posterior sampling accumulates at variable speed across the branches of the tree.
To further improve the sampling of our HMC algorithm, one might use an approximation of the posterior covariance of $\allRates$ for the mass matrix $\massMatrix$ to appropriately weight momentum updates in the HMC algorithm \citep{neal2011mcmc}.
Possible approximations include the Hessian of the log-posterior (a local approximation of the curvature of branch-rate multiplier space) or the sample variance across each dimension.
An important consideration in choosing an appropriate $\massMatrix$ is whether one is studying under a fixed or random phylogeny $\phylogeny$.
Since varying $\phylogeny$ in the posterior often creates multimodal distributions of $\allRates$, local approximations such as the Hessian may be of limited assistance in such cases.

We show in our application to mammalian life history that our computationally efficient HMC algorithm imbues the RRW model with the ability to handle large trees with thousands of taxa.
By applying the RRW model to this massive example we confirm that large mammals have `slower' life history characteristics, exhibited by the positive correlation between body mass, gestation length and weaning age, while smaller mammals scale in the opposite manner and tend to have high litter frequency and size, see Figure (\ref{fig:mammalCorrelation}).
The posterior mean correlation between litter frequency and litter size changes sign under each model, but with low posterior probability reflecting a lack of correlation between these traits.
Importantly, our method allows us to obtain results in 48 hours where the previous UMH method would take approximately 33 days.

In a time where biological data are more prolific than ever, scalable approaches to complex models of evolution such as the RRW prove increasingly useful in a variety of applications.
From spatial epidemiology where determining the dispersal rate of an infectious disease is crucial, to evolutionary ecology where understanding life-history can provide insight into declining animal populations, the analysis of data is becoming a bottleneck to the scientific process and the need for computationally faster approaches stands evident.
We hope that this work will serve to improve the speed of such analyses.


\section{Acknowledgments}

This work was supported by the European Research Council under the European Union's Horizon 2020 research and innovation programme (grant agreement number 725422 - ReservoirDOCS); The Artic Network from the Wellcome Trust (grant number 206298/Z/17/Z); National Science Foundation (grant number DMS 1264153); National Institutes of Health (grants R01 AI107034, U19 AI135995, and T32 GM008185 to AAF); and Research Foundation -- Flanders (`Fonds voor Wetenschappelijk Onderzoek -- Vlaanderen',  G066215N, G0D5117N and G0B9317N) to PL.


\section{Appendices}
\label{sec:computationalProcedures}
We provide public access to all BEAST XML files used in this work at \url{http://github.com/suchard-group/RRW_at_scale}.

\subsection{Pre-Order Partial Likelihood}
\label{sec:preOrderDerivation}
Here we derive a generalized version of the pre-order recursive algorithm proposed by \cite{cybis2015} to compute $\cDensity{\latentData_{\nodeIndexOne}}{\latentData_{\above{\nodeIndexOne}}}$ for all $\nodeIndexOne$ in $\order{\nTaxa}$.
We begin with the law of total probability,

\begin{equation}
	\cDensity{\latentData_{\nodeIndexOne} }{\latentData_{\above{\nodeIndexOne}}} \propto \int \cDensity{\latentData_{\nodeIndexOne}}{\latentData_{\nodeIndexThree}} \cDensity{\latentData_{\below{\nodeIndexTwo}}}{\latentData_{\nodeIndexThree}}
	\cDensity{\latentData_{\nodeIndexThree}}{\latentData_{\above{\nodeIndexThree}}} \ \dd \latentData_{\nodeIndexThree}
	\label{eq:preorder1}
\end{equation}
for node $\nodeIndexOne$ with parent $\nodeIndexThree$ and sibling $\nodeIndexTwo$. Recalling that

\begin{equation}
	\begin{aligned}
	\cDensity{\latentData_{\nodeIndexOne}}{\latentData_{\nodeIndexThree}} &= \mvnDensity{\latentData_{\nodeIndexOne}; \ \latentData_{\nodeIndexThree}}{
		\branchLength{\nodeIndexOne} \nodeVarianceFunction{\rate{\nodeIndexOne}}}, \  \text{and}\\
\cDensity{\latentData_{\below{\nodeIndexTwo}}}{\latentData_{\nodeIndexThree}}
&\propto
\mvnDensity{\latentData_{\nodeIndexThree}; \ \nodeMean{\nodeIndexTwo}}{(\nodePrecision{\nodeIndexTwo}^* )\inverse},
	\end{aligned}
\end{equation}
we identify Equation (\ref{eq:preorder1}) as a recursive expression whose solution has the form
\begin{equation}
	\cDensity{\latentData_{\nodeIndexOne}}{\latentData_{\above{\nodeIndexOne}}}
	=
	\mvnDensity{\latentData_{\nodeIndexOne}; \ \nodePreMean{\nodeIndexOne}}{\nodePrePrecision{\nodeIndexOne}},
\end{equation}
with presently undetermined pre-order mean $\nodePreMean{\nodeIndexOne}$ and pre-order precision $\nodePrePrecision{\nodeIndexOne}$.

We unravel these quantities by first identifying that $\cDensity{\latentData_{2 \nTaxa - 1}}{\latentData_{\above{2 \nTaxa - 1}}} = \density{\latentData_{2 \nTaxa - 1}}$ and set $\nodePreMean{2 \nTaxa - 1} = \rootTraitMean$ and $\nodePrePrecision{2 \nTaxa - 1} = \rootSampleSize \diffusionVariance \inverse$. Then proceeding in pre-order fashion for $\nodeIndexOne = 2 \nTaxa - 2, \ldots, 1$
\begin{equation}
	\begin{aligned}
	\nodePrePrecision{\nodeIndexOne}
	&=
	\Big( (\nodePrePrecision{\nodeIndexOne}^{*}) \inverse
	+ t_i \nodeVarianceFunction{\rate{\nodeIndexOne}} \Big)^{-1}
	\ \text{where}\\
	\nodePrePrecision{\nodeIndexOne}^{*}
	&=
	 \nodePrecision{\nodeIndexTwo}^* + \nodePrePrecision{\nodeIndexThree},
	\ \text{and}\\
	\nodePreMean{\nodeIndexOne}
	&=
	(\nodePrePrecision{\nodeIndexOne}^{*}) \inverse \big( \nodePrecision{\nodeIndexTwo}^* \nodeMean{\nodeIndexTwo}
	+
	\nodePrePrecision{\nodeIndexThree} \nodePreMean{\nodeIndexThree} \big).
	\end{aligned}
	\end{equation}

\subsection{Pseudo-inverse}
\label{sec:pseudoInverseExplanation}

The pseudo-inverse used in the post-order tree traversal and defined by \cite{bastide2018inference} and \cite{hassler2019inferring} is an operation for inverting precision and variance matrices with diagonal entries that take the value $\infty$.
To invert a diagonal precision matrix, $\nodePrecision{\nodeIndexOne}$ with entries $\infty$ and $0$, we define $\infty \specialInverse = 0$ and $0 \specialInverse = \infty$.
To invert the variance matrix, $ \left( \nodePrecision{\nodeIndexOne} \specialInverse + \branchLength{\nodeIndexOne} \observedMatrix{\nodeIndexOne} \nodeVarianceFunction{\rate{\nodeIndexOne}} \observedMatrix{\nodeIndexOne} \right)$ we invert the block matrix of observed trait covariation and invert the remaining diagonal elements using the convention that $\infty \specialInverse = 0$.


\begin{thebibliography}{}

\bibitem[Bastide {\em et~al.}(2018)Bastide, An{\'e}, Robin, and
  Mariadassou]{bastide2018inference}
Bastide, P., An{\'e}, C., Robin, S., and Mariadassou, M. (2018).
\newblock Inference of adaptive shifts for multivariate correlated traits.
\newblock {\em {Systematic Biology}\/}, {\bf 67}(4), 662--680.

\bibitem[Bedford {\em et~al.}(2014)Bedford, Suchard, Lemey, Dudas, Gregory,
  Hay, McCauley, Russell, Smith, and Rambaut]{bedford2014integrating}
Bedford, T., Suchard, M.~A., Lemey, P., Dudas, G., Gregory, V., Hay, A.~J.,
  McCauley, J.~W., Russell, C.~A., Smith, D.~J., and Rambaut, A. (2014).
\newblock Integrating influenza antigenic dynamics with molecular evolution.
\newblock {\em {eLife}\/}, {\bf 3}, e01914.

\bibitem[Betancourt(2017)Betancourt]{betancourt2017conceptual}
Betancourt, M. (2017).
\newblock {A conceptual introduction to Hamiltonian Monte Carlo}.
\newblock {\em arXiv preprint arXiv:1701.02434\/}.

\bibitem[Biek {\em et~al.}(2007)Biek, Henderson, Waller, Rupprecht, and
  Real]{biek2007high}
Biek, R., Henderson, J.~C., Waller, L.~A., Rupprecht, C.~E., and Real, L.~A.
  (2007).
\newblock A high-resolution genetic signature of demographic and spatial
  expansion in epizootic rabies virus.
\newblock {\em Proceedings of the {National Academy of Sciences}\/}, {\bf
  104}(19), 7993--7998.

\bibitem[Bryant {\em et~al.}(2005)Bryant, Galtier, and
  Poursat]{bryant2005likelihood}
Bryant, D., Galtier, N., and Poursat, M.-A. (2005).
\newblock Likelihood calculation in molecular phylogenetics.
\newblock In O.~Gascuel, editor, {\em Mathematics of Evolution and
  Phylogeny\/}, pages 33--62. Oxford Univ. Press.

\bibitem[Cybis {\em et~al.}(2015)Cybis, Sinsheimer, Bedford, Mather, Lemey, and
  Suchard]{cybis2015}
Cybis, G.~B., Sinsheimer, J.~S., Bedford, T., Mather, A.~E., Lemey, P., and
  Suchard, M.~A. (2015).
\newblock Assessing phenotypic correlation through the multivariate
  phylogenetic latent liability model.
\newblock {\em The Annals of Applied Statistics\/}, {\bf 9}(2), 969.

\bibitem[de~Silva and Leimgruber(2019)de~Silva and
  Leimgruber]{de2019demographic}
de~Silva, S. and Leimgruber, P. (2019).
\newblock Demographic tipping points as early indicators of vulnerability for
  slow-breeding megafaunal populations.
\newblock {\em {Frontiers in Ecology and Evolution}\/}, {\bf 7}, 171.

\bibitem[Faria {\em et~al.}(2014)Faria, Rambaut, Suchard, Baele, Bedford, Ward,
  Tatem, Sousa, Arinaminpathy, P{\'e}pin, {\em et~al.}]{faria2014early}
Faria, N.~R., Rambaut, A., Suchard, M.~A., Baele, G., Bedford, T., Ward, M.~J.,
  Tatem, A.~J., Sousa, J.~D., Arinaminpathy, N., P{\'e}pin, J., {\em et~al.}
  (2014).
\newblock The early spread and epidemic ignition of {HIV-1} in human
  populations.
\newblock {\em Science\/}, {\bf 346}(6205), 56--61.

\bibitem[Felsenstein(1981)Felsenstein]{felsenstein1981}
Felsenstein, J. (1981).
\newblock Evolutionary trees from {DNA} sequences: a maximum likelihood
  approach.
\newblock {\em Journal of Molecular Evolution\/}, {\bf 17}(6), 368--376.

\bibitem[Felsenstein(1985)Felsenstein]{felsenstein1985}
Felsenstein, J. (1985).
\newblock Phylogenies and the comparative method.
\newblock {\em The American Naturalist\/}, {\bf 125}(1), 1--15.

\bibitem[Fritz {\em et~al.}(2009)Fritz, Bininda-Emonds, and
  Purvis]{fritz2009geographical}
Fritz, S.~A., Bininda-Emonds, O.~R., and Purvis, A. (2009).
\newblock Geographical variation in predictors of mammalian extinction risk:
  big is bad, but only in the tropics.
\newblock {\em Ecology Letters\/}, {\bf 12}(6), 538--549.

\bibitem[Gray {\em et~al.}(2010)Gray, Veras, Santos, and
  Salemi]{gray2010evolutionary}
Gray, R., Veras, N., Santos, L., and Salemi, M. (2010).
\newblock Evolutionary characterization of the {West} {Nile} virus complete
  genome.
\newblock {\em Molecular Phylogenetics and Evolution\/}, {\bf 56}(1), 195--200.

\bibitem[Hassler {\em et~al.}(2019)Hassler, Tolkoff, Allen, Ho, Lemey, and
  Suchard]{hassler2019inferring}
Hassler, G., Tolkoff, M.~R., Allen, W.~L., Ho, L. S.~T., Lemey, P., and
  Suchard, M.~A. (2019).
\newblock Inferring phenotypic trait evolution on large trees with many
  incomplete measurements.
\newblock {\em arXiv preprint arXiv:1906.03222\/}.

\bibitem[Hastings(1970)Hastings]{hastings1970monte}
Hastings, W.~K. (1970).
\newblock Monte Carlo sampling methods using {Markov} chains and their
  applications.
\newblock {\em Biometrika\/}, {\bf 57}(1), 97 -- 109.

\bibitem[Ives and Garland~Jr(2009)Ives and Garland~Jr]{ives2009phylogenetic}
Ives, A.~R. and Garland~Jr, T. (2009).
\newblock Phylogenetic logistic regression for binary dependent variables.
\newblock {\em Systematic Biology\/}, {\bf 59}(1), 9--26.

\bibitem[Jones {\em et~al.}(2009)Jones, Bielby, Cardillo, Fritz, O'Dell, Orme,
  Safi, Sechrest, Boakes, Carbone, {\em et~al.}]{jones2009pantheria}
Jones, K.~E., Bielby, J., Cardillo, M., Fritz, S.~A., O'Dell, J., Orme, C.
  D.~L., Safi, K., Sechrest, W., Boakes, E.~H., Carbone, C., {\em et~al.}
  (2009).
\newblock {PanTHERIA}: a species-level database of life history, ecology, and
  geography of extant and recently extinct mammals.
\newblock {\em Ecology\/}, {\bf 90}(9), 2648--2648.

\bibitem[Lemey {\em et~al.}(2010)Lemey, Rambaut, Welch, and Suchard]{lemey2010}
Lemey, P., Rambaut, A., Welch, J.~J., and Suchard, M.~A. (2010).
\newblock Phylogeography takes a relaxed random walk in continuous space and
  time.
\newblock {\em Molecular Biology and Evolution\/}, {\bf 27}(8), 1877--1885.

\bibitem[Levine and Casella(2006)Levine and Casella]{levine2006}
Levine, R.~A. and Casella, G. (2006).
\newblock Optimizing random scan {Gibbs} samplers.
\newblock {\em Journal of Multivariate Analysis\/}, {\bf 97}(10), 2071--2100.

\bibitem[Liu(2008)Liu]{liu2008monte}
Liu, J.~S. (2008).
\newblock {\em Monte {Carlo} strategies in scientific computing\/}.
\newblock Springer Science \& Business Media.

\bibitem[Millar and Zammuto(1983)Millar and Zammuto]{millar1983life}
Millar, J.~S. and Zammuto, R.~M. (1983).
\newblock Life histories of mammals: an analysis of life tables.
\newblock {\em {Ecology}\/}, {\bf 64}(4), 631--635.

\bibitem[Neal(2011)Neal]{neal2011mcmc}
Neal, R.~M. (2011).
\newblock {MCMC} using {Hamiltonian} dynamics.
\newblock In S.~Brooks, A.~Gelman, G.~L. Jones, and X.-L. Meng, editors, {\em
  Handbook of Markov Chain Monte Carlo\/}, volume~2. CRC Press New York, NY.

\bibitem[Oli(2004)Oli]{oli2004fast}
Oli, M.~K. (2004).
\newblock The fast--slow continuum and mammalian life-history patterns: an
  empirical evaluation.
\newblock {\em {Basic and Applied Ecology}\/}, {\bf 5}(5), 449--463.

\bibitem[Pacifici {\em et~al.}(2017)Pacifici, Visconti, Butchart, Watson,
  Cassola, and Rondinini]{pacifici2017species}
Pacifici, M., Visconti, P., Butchart, S.~H., Watson, J.~E., Cassola, F.~M., and
  Rondinini, C. (2017).
\newblock {Species' traits influenced their response to recent climate change}.
\newblock {\em {Nature Climate Change}\/}, {\bf 7}(3), 205.

\bibitem[Petersen {\em et~al.}(2008)Petersen, Pedersen, {\em
  et~al.}]{matrixcookbook}
Petersen, K.~B., Pedersen, M.~S., {\em et~al.} (2008).
\newblock The matrix cookbook.
\newblock {\em Technical University of Denmark\/}, {\bf 7}(15), 510.

\bibitem[Petersen {\em et~al.}(2013)Petersen, Brault, and
  Nasci]{petersen2013west}
Petersen, L.~R., Brault, A.~C., and Nasci, R.~S. (2013).
\newblock West {N}ile virus: review of the literature.
\newblock {\em Journal of the American Medical Association\/}, {\bf 310}(3),
  308--315.

\bibitem[Pybus {\em et~al.}(2012)Pybus, Suchard, Lemey, Bernardin, Rambaut,
  Crawford, Gray, Arinaminpathy, Stramer, Busch, {\em
  et~al.}]{pybus2012unifying}
Pybus, O.~G., Suchard, M.~A., Lemey, P., Bernardin, F.~J., Rambaut, A.,
  Crawford, F.~W., Gray, R.~R., Arinaminpathy, N., Stramer, S.~L., Busch,
  M.~P., {\em et~al.} (2012).
\newblock Unifying the spatial epidemiology and molecular evolution of emerging
  epidemics.
\newblock {\em Proceedings of the National Academy of Sciences\/}, {\bf
  109}(37), 15066--15071.

\bibitem[Santini {\em et~al.}(2016)Santini, Cornulier, Bullock, Palmer, White,
  Hodgson, Bocedi, and Travis]{santini2016trait}
Santini, L., Cornulier, T., Bullock, J.~M., Palmer, S.~C., White, S.~M.,
  Hodgson, J.~A., Bocedi, G., and Travis, J.~M. (2016).
\newblock A trait-based approach for predicting species responses to
  environmental change from sparse data: how well might terrestrial mammals
  track climate change?
\newblock {\em {Global Change Biology}\/}, {\bf 22}(7), 2415--2424.

\bibitem[Schluter {\em et~al.}(1997)Schluter, Price, Mooers, and
  Ludwig]{schluter1997likelihood}
Schluter, D., Price, T., Mooers, A.~{\O}., and Ludwig, D. (1997).
\newblock Likelihood of ancestor states in adaptive radiation.
\newblock {\em Evolution\/}, {\bf 51}(6), 1699--1711.

\bibitem[Snapinn {\em et~al.}(2007)Snapinn, Holmes, Young, Bernard, Kramer, and
  Ebel]{snapinn2007declining}
Snapinn, K.~W., Holmes, E.~C., Young, D.~S., Bernard, K.~A., Kramer, L.~D., and
  Ebel, G.~D. (2007).
\newblock Declining growth rate of {West} {Nile} virus in {North} {America}.
\newblock {\em Journal of Virology\/}, {\bf 81}(5), 2531--2534.

\bibitem[Stearns(2000)Stearns]{stearns2000life}
Stearns, S.~C. (2000).
\newblock Life history evolution: successes, limitations, and prospects.
\newblock {\em Naturwissenschaften\/}, {\bf 87}(11), 476--486.

\bibitem[Suchard {\em et~al.}(2018)Suchard, Lemey, Baele, Ayres, Drummond, and
  Rambaut]{suchard2018}
Suchard, M.~A., Lemey, P., Baele, G., Ayres, D.~L., Drummond, A.~J., and
  Rambaut, A. (2018).
\newblock Bayesian phylogenetic and phylodynamic data integration using {BEAST}
  1.10.
\newblock {\em Virus Evolution\/}, {\bf 4}(1), vey016.

\end{thebibliography}
\end{document}